\documentstyle[aps]{revtex}

\begin{document}
\author{O. G. Balev$^{a,b}$ and Nelson Studart$^{a}$}
\address{$^{a}$Departmento de Fisica, Universidade Federal de S\~{a}o Carlos,\\
13565-905, S\~{a}o Carlos, S\~{a}o Paulo, Brazil \\
$^{b}$Institute of Semiconductor Physics, National Academy of Sciences,\\
45 Pr. Nauky, Kiev 252650, Ukraine}
\date{August 12, 1999}
\title{Temperature Effects on Edge Magnetoplasmons in the Quantum Hall\\
Regime}
\maketitle

\begin{abstract}
A {\it microscopic} treatment of edge magnetoplasmons (EMPs) is presented
for the case of not-too-low temperatures in which the inequality $k_{B}T\gg
\hbar v_{g}/\ell _{0}$, where $v_{g}$ is the group velocity of the edge
states and $\ell _{0}$ is the magnetic length, is fulfilled, and for filling
factors $\nu =1(2)$. We have obtained independent EMP modes spatially
symmetric and antisymmetric with respect to the edge. We describe in detail
the spatial structure and dispersion relations of the new edge waves (edge
helicons, dipole, quadrupole and octupole EMPs), which have the
characteristic length $\ell _{T}=\ell _{0}^{2}k_{B}T/\hbar v_{g}$. We have
found that, in contrast to well-known results for a spatially homogeneous
dissipation within the channel, the damping of the fundamental EMP at
not-too-low temperatures is not quantized and has a $T^{-1}$ dependence.

PACS\ \ 73.20.Dx, 73.40.Hm
\end{abstract}

\section{INTRODUCTION.}

Edge magnetoplasmons (EMPs) in the two-dimensional electron system (2DES)
have received much attention in recent years. Experimental studies have been
performed to determine the dispersion relation and the role of edge states
in the transport properties of the 2DES both on liquid helium\cite
{mast85,glattli85,kirichek95} as in high-mobility AlGaAs-GaAs
heterostructures \cite
{talyanskii89,wassermeier90,grodnensky91,ashoori92,talyanskii92,zhitenev93,ernst96,balaban97}%
. The interest have even increased with the advent of time-resolved
transport experiments \cite{ashoori92,zhitenev93,ernst96}. From a
theoretical point of view, a lot of work have been also devoted to study the
characteristics of these collective excitations propagating along the edge
of the 2DES in the presence of a normal magnetic field $B$ and different
edge-wave mechanisms have been proposed \cite
{fetter86,volkov88,wen91,stone91,aleiner94,chamon94,giovanazzi94,han97,zulicke97,balev97,balev98,balev99}%
. First, the EMPs dispersion were determined theoretically within
essentially classical models \cite{volkov88,aleiner94} in which the charge
density varies at the edge, but the edge position of the 2DES is kept
constant. Other, distinctly different, quantum-mechanical edge-wave
mechanisms was proposed \cite{wen91,stone91,chamon94,giovanazzi94} in which
only the edge change and the density profile is taken as the unperturbed
2DES with respect to the fluctuating edge.

Recently a microscopic model was proposed in Refs.\cite{balev97,balev98}
that effectively incorporates the edge-wave mechanisms mentioned above in
the quantum Hall effect (QHE) regime. Even though EMPs have been studied in
the limit of low temperatures $k_{B}T\ll \hbar v_{g}/\ell _{0}$, where $%
v_{g} $ is the group velocity of the edge states and the magnetic length $%
\ell _{0}=\sqrt{\hbar /m^{*}\omega _{c}}$ with $\omega _{c}=|e|B/m^{*}c$, in
the calculation of the current density ${\bf J}$ was assumed that the
components of the electric field ${\bf E}$ of the wave are smooth on the $%
\ell _{0}$ scale. However, this assumption can not be well justified for
EMPs at very low temperatures. Here, we extend the approach of Refs. \cite
{balev97,balev98} for not-too-low temperatures, where $\hbar \omega _{c} \gg
k_{B}T\gg \hbar v_{g}/\ell _{0}$. In this regime the typical scales of
in-plane components of ${\bf E}$ are of the order of $\ell_{T}=\ell_{0}^{2}
k_{B}T/\hbar v_{g}$, which is much larger than $\ell_{0}$.

Our model to treat EMPs consists in considering a 2DES, of width $W$, length 
$L_{x}=L$, and negligible thickness, in the presence of a strong $B$
parallel to the $z$ axis, such that only the $n=0$ Landau level (LL) is
occupied. The 2DES is confined along the $y$ axis by a parabolic potential
at the edges given by $V_{y}^{^{\prime }}=0$, for $y_{l}<y<y_{r}$, $%
V_{y}^{^{\prime }}=m^{*}\Omega ^{2}(y-y_{r})^{2}/2$ for $y>y_{r}>0$, and $%
V_{y}^{^{\prime }}=m^{*}\Omega ^{2}(y-y_{l})^{2}/2$ for $y<y_{l}<0$. We
assume that the confinement is smooth on the scale of $\ell _{0}$ such that $%
\Omega \ll \omega _{c}$ and $|k_{x}|W\gg 1$ such that it is reasonable to
consider the EMP along the right edge of the channel, in the form $A(\omega
,k_{x},y)\exp [-i(\omega t-k_{x}x)]$, totally independent of the left edge.
We consider the QHE regime, at filling factors $\nu =1$ or $\nu =2$ in
samples with areal dimensions sufficiently large, as it is typical in EMP
experiments. So, the inter-edge electron transitions and the inter-edge
Coulomb interaction can be neglected. At $\nu =1$, we assume that the
spin-splitting, caused by many-body effects, is strong enough to neglect the
contribution from the upper spin-split LL. At $\nu =2$, we neglect
spin-splitting.

There are at least two essential aspects to take into account in order to
determine the dispersion relation as well the spatial structures of the
EMPs, One is the role of dissipation even in the QHE regime. Recently it was
shown, for sufficiently smooth confinement, that the dissipation comes from
the intralevel-intraedge electronic transitions due to scattering by
piezoelectric phonons and occurs mainly near the edges of the channel \cite
{balev93}. In the linear response regime, this is the main dissipation
mechanism if $v_{g}>s$, the speed of sound, {\it i.e}., for channels of
width $W\alt 100\ \mu $m and $T\alt 1$ K. The dissipation in the bulk is
exponentially suppressed for $\hbar \omega _{c}/k_{B}T\gg 1$. So, the
properties of the EMPs must be strongly modified when the dissipation is
localized near the edges in comparison with previous works in which the
dissipation occurs homogeneously over the channel width \cite{volkov88}.

The second point is the profile of the unperturbed electron density $n_{0}(y)
$ across the edge. It was shown that the introduction of a smooth $n_{0}(y)$%
, considered in Ref. \cite{aleiner94}, leads to the appearance of new
acoustic modes in addition to those found in Ref. \cite{volkov88} in which
the electron density drops abruptly at the edge. In Fig. 1, we compare our
calculated unperturbed density profile $n_{0}(y)$ with the Volkov-Mikhailov%
\cite{volkov88} and Aleiner-Glazman\cite{aleiner94} models. Notice that $%
n_{0}(y)$ is presented on the scale of $\ell _{T}\gg \ell _{0}$ near the
edge instead of $\ell _{0}$, as in Fig. 1 of Ref. \cite{balev98}. Our
calculated exact density profile $n_{0}(y)/n_{0}$ (solid curve), where $n_{0}
$ is the bulk value, is shown together with that of Refs. \cite{volkov88}
(short-dashed curve) and \cite{aleiner94} (dotted curve). The dash-dotted
curve represents our approximate density profile given by $%
n_{0}(y)/n_{0}=[1-\tanh (Y/2)]/2$, where $Y=(y-y_{r0})/\ell _{T}$. As we can
see, our analytical profile is very close to the exact one in the actual
region if the condition $k_{e}\gg k_{B}T/\hbar v_{g}\gg \ell _{0}^{-1}$ is
fulfilled, where the characteristics edge wave number $k_{e}=(\omega
_{c}/\hbar \Omega )\sqrt{2m^{*}\Delta _{F}}$, with $\Delta _{F}$ being the
Fermi energy measured from the bottom of the $n=0$ LL, or $\Delta
_{F}=E_{F0}-\hbar \omega _{c}/2$. The density profile in the Aleiner-Glazman
model is obtained taking $n_{0}(y)/n_{0}=(2/\pi )\arctan \sqrt{(y_{re}-y)/a}$
and $a/\ell _{0}=20$ which corresponds approximately to $a=2000$ \AA . As it
can be seen, our density profiles are very different from the other ones.
For the calculation of the solid curve, the chosen parameters, related to
GaAs-based heterostructures, are $B=5.9$ T, $m^{*}=0.067m_{0}$, $\omega
_{c}/\Omega =30$, $T=18$ K, $\ell _{T}/\ell _{0}=5,$ and taking $\Delta
_{F}=\hbar \omega _{c}/2$ such that $v_{g}=\Omega \ell _{0}\approx 5\times
10^{5}$ cm/sec.

We will show that the combination of our density profile and the fact that
dissipation is localized near the edge leads to strong modifications of the
EMP behavior. These changes as well as the new EMPs resulting from the
present microscopic approach in the regime of not-too-low temperatures are
the subject of this work.

The organization of the paper is as follows. In Sec. II, for completeness,
we present the expressions for the inhomogeneous current densities and
conductivities for not-too-low temperatures and in the quasi-static regime
following the treatment given in Ref. \cite{balev96}. We also obtain the
general equation for the EMPs with dissipation at the edges. In contrast
with EMPs at very low $T$, we neglect the nonlocal contributions to the
current density\cite{balev97,balev98} since their spatial structure is
smooth in the scale of $\ell _{0}$. Further, in Sec. III, we elaborate on
the integral equations for symmetric and antisymmetric EMPs and devise a
method to solve them. In Sec.IV we derive the dispersion relations and
charge-density amplitudes of the symmetric and antisymmetric EMPs and
describe in detail the new edge waves. Finally, in Sec. V, we compare our
theory with experiment and make concluding remarks.

\section{INTEGRAL EQUATION FOR EMPS WITH DISSIPATION AT THE EDGES}

In the low-frequency limit ($\omega \ll \omega _{c})$ for the EMP, the
current density can be calculated in the quasi-static approximation and
using the fact that wavelength $\lambda \gg \ell _{0} \alt 10^{-6}$cm and
the characteristic scale along $y$ is typically of the order of $\ell _{T}
\gg \ell _{0}$ for not-too-low $T$ (this condition can be broken only for
very high multi-pole modes). Hence the results of Ref. \cite{balev96} for
the components of the current density can be also considered here and are
given by

\begin{equation}
j_{y}(y)=\sigma _{yy}(y)E_{y}(y)+\sigma _{yx}^{0}(y)E_{x}(y),  \label{1} \\
\end{equation}

\begin{equation}
j_{x}(y)=\sigma _{xx}(y)E_{x}(y)-\sigma _{yx}^{0}(y)E_{y}(y)+v_{g}\rho
(\omega ,k_{x},y).  \label{2}
\end{equation}
We have suppressed the exponential factor $\exp [-i(\omega t-k_{x}x)]$
common to all terms in Eqs. (\ref{1}) and (\ref{2}) and obviously $E_{\mu
}(y)$ depends also on $\omega $ and $k_{x}$. As shown in Refs. \cite{balev93}
and \cite{balev96}, $\sigma _{yy}(y)$ is strongly localized with exponential
decay at the edge, within a distance $\alt \ell _{T}$ from it, for $\hbar
\omega _{c}\gg k_{B}T\gg \hbar v_{g}/\ell _{0}$. The last term on the
right-hand side (RHS) of Eq. (\ref{2}), absent in Ref. \cite{balev96},
represents the contribution of the current density along $x$, associated
with the advection of a wave distortion $\rho (\omega ,k_{x},y)$ of the
charge localized near the edge. For $\nu =1$ we have

\begin{equation}
\sigma _{yx}^{0}(y)=\frac{e^{2}}{2\pi \hbar }\int_{-\infty }^{\infty
}dy_{0\alpha }f_{\alpha 0}\Psi _{0}^{2}(y-y_{0\alpha }),  \label{3}
\end{equation}
where $\alpha \equiv \{0,k_{x\alpha }\}$,\ $y_{0\alpha }=\ell
_{0}^{2}k_{x\alpha }$, $\Psi _{n}(y)$ is the harmonic-oscillator function,
and $f_{\alpha 0} \equiv f_{0}(k_{x\alpha })=1/[1+\exp ((E_{\alpha
0}-E_{F0})/k_{B}T)]$ is the Fermi-Dirac function. $E_{F0}$ is the Fermi
level measured from the bottom of the lowest electric subband. Considering
only the flat part of the confining potential, $y_{l}\leq y_{0\alpha} \leq
y_{r}$, we have $E_{\alpha 0}=\hbar \omega _{c}/2$. For the right edge
region, $y_{0\alpha }\geq y_{r}$, we obtain

\begin{equation}
E_{\alpha 0}\equiv E_{0}(y_{0\alpha })=\hbar \omega _{c}/2+m^{*}
\Omega^{2}(y_{0\alpha }-y_{r})^{2}/2.  \label{4}
\end{equation}

In our regime of interest, $\ell _{T}\gg \ell _{0}$, taking into account
that the typical $y$-scale of $f_{\alpha 0}$ and $\Psi
_{0}^{2}(y-y_{0\alpha})$ are $\ell_{T}$ and $\ell_{0}$ respectively, we
evaluate the integral in the RHS of Eq. (\ref{3}) to obtain

\begin{equation}
\sigma _{yx}^{0}(y)\approx \frac{e^{2}}{2\pi \hbar }\frac{1}{%
[1+exp((E_{0}(y))-E_{F0})/k_{B}T)]}.  \label{5}
\end{equation}
At the edge of the $n=0$ LL, $y=y_{re}=\ell _{0}^{2}k_{re}$, where $%
f_{0}(k_{re})=1/2$, we have $\sigma _{yx}^{0}(y_{re})=e^{2}/(4\pi \hbar )$.
So $\sigma _{yx}^{0}(y)$ decreases on the scale of $\ell _{T}$ near the edge
and behaves like the density profile shown by the solid curve in Fig. 1.

We consider only the electron-phonon interaction, since it is the most
essential process for the required conditions \cite{balev93}. Because the
dependence of $E_{x}(y)$ on the $\ell _{0}$ scale is quite smooth, we can
assume that $\sigma _{xx}(y)$ can be approximated by $\sigma _{yy}(y)$ and
following closely the results of Ref. \cite{balev96}, we can write

\begin{eqnarray}
\sigma _{yy}(y) &=&\frac{\pi e^{2}\ell _{0}^{4}}{4\hbar Lk_{B}T}
\sum_{k_{x\alpha }{\bf q}}|C_{{\bf q}}|^{2}q_{x}^{2}
[f_{0}(k_{x\alpha}-q_{x})-f_{0}(k_{x\alpha })] \delta [E_{0}(k_{x\alpha })-
E_{0}(k_{x\alpha}-q_{x})-\hbar \omega _{\vec{q}}]  \nonumber \\
&&\times e^{-(q_{x}^{2}+q_{y}^{2})\ell _{0}^{2}/2}\ \sinh ^{-2}(\frac{\hbar
\omega _{\vec{q}}}{2k_{B}T})\ [\Psi _{0}^{2}(y-y_{0}(k_{x\alpha}-q_{x}))+
\Psi _{0}^{2}(y-y_{0}(k_{x\alpha }))].  \label{6}
\end{eqnarray}
For temperatures in the QHE regime, the relevant contributions arise only
from acoustical (DA) or piezoelectrical (PA) phonons with dispersion $\omega
_{{\bf q}}=sq$, where $q=\sqrt{q_{x}^{2}+q_{y}^{2}+q_{z}^{2}}$. Then the
interaction strength $|C_{{\bf q}}|^{2}=(c^{^{\prime
}}/L_{x}L_{y}L_{z})q^{\pm 1}$, where $+1$ is for DA and $-1$ for PA-phonons
respectively, and $c^{\prime }$ is the electron-phonon coupling constant.

Using the Eqs. (\ref{1}), (\ref{2}), (\ref{5}), (\ref{6}), the Poisson
equation and the linearized continuity equation, we obtain the integral
equation for $\rho (\omega ,k_{x},y)$

\begin{eqnarray}
-i(\omega -k_{x}v_{g})&&\rho (\omega ,k_{x},y)+ \frac{2}{\epsilon}%
\{k_{x}^{2}\sigma _{xx}(y)- ik_{x}\frac{d}{dy}[\sigma _{yx}^{0}(y)] 
\nonumber \\
* &&-\sigma _{yy}(y)\frac{d^{2}}{dy^{2}}- \frac{d}{dy}[\sigma _{yy}(y)]\frac{%
d}{dy}\} \int_{-\infty }^{\infty }dy^{\prime }K_{0}(|k_{x}||y-y^{\prime }|)
\rho(\omega ,k_{x},y^{\prime })=0,  \label{7}
\end{eqnarray}
where $K_{0}(x)$ is the modified Bessel function. We can see from Eqs. (\ref
{5}) and (\ref{6}) that $\sigma _{yy}(y)$ and $d\sigma _{yx}^{0}(y)/dy$, as
well $\sigma _{xx}(y)$, are exponentially localized within a distance of
order of $\ell _{T}$ from the right edge at $y_{re}=y_{r}+\Delta y_{r}$
where $\Delta y_{r}=\ell _{0}^{2}k_{e}$ for $\hbar v_{g}\ll \ell _{0}k_{B}T$
and $W=2y_{re}$. For $k_{x\alpha }\equiv k_{re}=y_{r}/\ell _{0}^{2}+k_{e}$, $%
f_{0}(k_{re})=1/2$, we have

\begin{equation}
v_{g}=\frac{1}{\hbar }\frac{\partial E_{0}(k_{re})}{\partial k_{x\alpha }}=%
\frac{\hbar \Omega ^{2}k_{e}}{m^{*}\omega _{c}^{2}}=\sqrt{\frac{2\Delta _{F}%
}{m^{*}}}\frac{\Omega }{\omega _{c}}.  \label{8}
\end{equation}
or alternatively $v_{g}=cE_{e}/B$, where $E_{e}=\Omega \sqrt{2m^{*}\Delta
_{F}}/|e|$ is the electric field describing the influence of the confining
potential. For a dissipationless classical 2DES we have, for finite $\omega $%
, $\sigma _{yy}(y)=\sigma _{xx}(y)=ie^{2}n_{0}(y)\omega /
m^{*}(\omega^{2}-\omega _{c}^{2})$ and $\sigma
_{yx}^{0}(y)=-e^{2}n_{0}(y)\omega_{c} /m^{*}(\omega ^{2}-\omega _{c}^{2})$,
where $n_{0}(y)$ is the electron density. In this case, Eq. (\ref{7}) is the
same as Eq. (4) of Ref. \cite{aleiner94}.

Equation (\ref{7}) is valid for a 2DES in the absence of metallic gates.
Experimentally, a metallic gate is sometimes placed on the top of the sample
at a distance $d$ from the 2DES \cite{zhitenev93}. For the gated sample, the
kernel $K_{0}$ in Eq. (\ref{7}) is replaced by $R_{g}=K_{0}(|k_{x}||y-y^{%
\prime }|)- K_{0}(|k_{x}|\sqrt{(y-y^{\prime })^{2}+4d^{2}})$ and if we have
air in the top of the 2DES, then $K_{0}$ is replaced by $%
R_{a}=K_{0}(|k_{x}||y-y^{\prime }|)+[(\epsilon -1)/(\epsilon
+1)]K_{0}(|k_{x}|\sqrt{(y-y^{\prime })^{2}+4d^{2}})$ \cite{balev97}. For
definiteness, we take the background dielectric constant $\epsilon$ to be
spatially homogeneous.

\section{TEMPERATURE EFFECTS ON DISPERSION EQUATIONS FOR EMPs}

At temperatures such that the inequality $k_{e} \gg k_{B}T/\hbar v_{g} \gg
\ell _{0}^{-1}$ is satisfied, we obtain, from Eq. (\ref{5}), that $%
d\sigma_{yx}^{0}(y)/dy= (e^{2}/2\pi \hbar )R_{0}(y)$, where the function $%
R_{0}(y)=-d[1+exp((E_{0}(y))-E_{F0})/k_{B}T)]^{-1}/dy \approx
(4\ell_{T})^{-1} \cosh ^{-2}(\bar{y}/2\ell _{T})]$, with $\bar{y}=y-y_{re}$,
is exponentially localized around $y_{re}$ within a distance of the order of 
$\ell_{T}$. In addition, from Eq. (\ref{6}), it follows that the dissipative
components of the conductivity tensor $\sigma _{\mu \mu }(y)$ are
proportional to $R_{0}(y)\equiv R_{0}(\bar{y})$, where $\mu$ stands for $x$
or $y$, and are also strongly concentrated nearby $y_{re}$. Notice also that
the condition $k_{e}\ell _{0}\gg 1$ also holds, due to the fact that $%
\omega_{c}/\Omega \gg 1$ if $\Delta _{F}\approx \hbar \omega _{c}/2$. Taking
into account that only the interaction with PA-phonons is relevant for $%
v_{g} \geq s$, from Eq. (\ref{6}) for $\nu =1$ we obtain

\begin{equation}
\tilde{\sigma}_{\mu \mu }=\sigma _{\mu \mu }(y)/R_{0}(\bar{y})= \frac{%
e^{2}\ell_{0}^{2}c^{^{\prime }}k_{B}T}{4\pi ^{2}\hbar ^{4}v_{g}^{3}},
\label{9}
\end{equation}
where, for GaAs-based heterostructures, $c^{\prime }=\hbar
(eh_{14})^{2}/2\rho _{V}s$, with $h_{14}=1.2\times 10^{7}$ V/cm, $\rho
_{V}=5.31$ gm/cm$^{3}$, and $s=2.5\times 10^{5}$ cm/sec. Hereafter in
numerical estimates and figures we are using these parameters.

One can see also, from Eq. (\ref{7}), that $\rho (\omega ,k_{x},y)\equiv
\rho (\omega ,k_{x},\bar{y})$ is also concentrated near the edge of the $n=0$
LL within a region of the order of $\ell _{T}$. Hereafter we use the
dimensionless variable $Y=\bar{y}/\ell _{T}$. Furthermore, Eq. (\ref{7}) is
invariant with respect to change $\bar{y}\rightarrow -\bar{y}$ for even EMP
modes, i.e. for which $\rho (\omega ,k_{x},\bar{y})=\rho (\omega ,k_{x},-%
\bar{y})$, or for odd EMP modes, $\rho (\omega ,k_{x},\bar{y})=-\rho (\omega
,k_{x},-\bar{y})$. It means that spatially symmetric (even) and spatially
antisymmetric (odd) EMP modes of Eq. (\ref{7}), with respect to $n=0$ LL
edge $y_{re}$, are totally independent from each other. 
As we will see, it is reasonable to look for a solution of Eq. (\ref{7}) in
a expansion series in terms of Laguerre polynomials $L_{n}(Y)$ in the
orthogonality interval $0\leq Y\leq \infty $. We define $\tilde{R}%
_{0}(Y)\equiv \exp (Y)*R_{0}(Y)=[\exp (Y)/4\ell _{T}]\cosh ^{-2}(Y/2)$ and
we can see that $\tilde{R}_{0}(Y)$ corresponds to the unperturbed electron
density profile, which we approximate it by $n_{0}(Y)/n_{0}=[1-\tanh (Y/2)]/2
$, which as we have seen practically coincide with the exact one, given by $%
2\pi \hbar \sigma _{yx}^{0}(y)/e^{2}$, where $\sigma _{yx}^{0}(y)$ is given
by Eq. (\ref{5}), if the conditions $k_{e}\gg k_{B}T/\hbar v_{g}\gg \ell
_{0}^{-1}$ are fulfilled. Then, from Eq. (\ref{7}), we obtain the integral
equation for symmetric EMPs for $Y\geq 0$ as

\begin{eqnarray}
(\omega -k_{x}v_{g}) &&\rho _{s}(\omega ,k_{x},\bar{y})-\frac{2}{\epsilon }%
\{(k_{x}\sigma _{yx}^{0}-ik_{x}^{2}\tilde{\sigma}_{xx})R_{0}(\bar{y})+i%
\tilde{\sigma}_{yy}\frac{d}{d\bar{y}}[R_{0}(\bar{y})\frac{d}{d\bar{y}}]\} 
\nonumber \\
&&*\   \nonumber \\
&&\int_{0}^{\infty }d\bar{y}^{\prime }[K_{0}(|k_{x}||\bar{y}-\bar{y}^{\prime
}|)+K_{0}(|k_{x}||\bar{y}+\bar{y}^{\prime }|)]\rho _{s}(\omega ,k_{x},\bar{y}%
^{\prime })=0,  \label{10}
\end{eqnarray}

In the long-wavelength limit, $k_{x}\ell _{T}\ll 1$, and for not too strong
dissipation at least one EMP mode must have a spatial behavior proportional
to $R_{0}(Y)$ and for large $Y$, $\rho _{s}(\omega ,k_{x},\bar{y})$ also
looks like $R_{0}(Y)$. Then, we write the solution of Eq.(\ref{10}) for the
symmetric EMPs for $\bar{y}\geq 0$ as

\begin{equation}
\rho _{s}(\omega ,k_{x},\bar{y})=\tilde{R}_{0}(Y)e^{-Y}\sum_{n=0}^{\infty
}\rho _{s}^{(n)}(\omega ,k_{x})L_{n}(Y).  \label{11}
\end{equation}
For $\bar{y}\leq 0,$ the expression for $\rho _{s}(\omega ,k_{x},\bar{y})$
follows trivially from Eq. (\ref{11}), just using $|Y|$ in the RHS of Eq. (%
\ref{11}). Notice that this expansion is valid only when the lowest LL is
occupied. We point out that $\tilde{R}_{0}(Y)$ is a rather weak dependence
on $Y$, especially for $Y\geq 1$, and tends to $1/4\ell _{T}$ as $%
Y\rightarrow \infty $.

In order to obtain the dispersion equation for the symmetric modes, we now
multiply Eq. (\ref{10}) by $L_{m}(Y)\tilde{R}_{0}^{-1}(Y)$ and integrate
over $Y=\bar{y}/\ell _{T}$ from $0$ to $\infty $. Then, taking into account
the Eq. (\ref{11}), we obtain

\begin{equation}
(\omega -k_{x}v_{g})\rho _{s}^{(m)}(\omega ,k_{x})-\sum_{n=0}^{\infty
}[S\;r_{mn}^{s}(k_{x})+S^{\prime }\;g_{mn}^{s}(k_{x})] \rho
_{s}^{(n)}(\omega,k_{x})=0,  \label{15}
\end{equation}
where, by assuming $\nu =1$, $S=(2/\epsilon )(k_{x}\sigma
_{yx}^{0}-ik_{x}^{2}\tilde{\sigma}_{xx})$, with $\sigma _{yx}^{0}=e^{2}/2\pi
\hbar $, $S^{\prime }=-2i\tilde{\sigma}_{yy}/\epsilon \ell _{T}^{2}$,

\begin{equation}
r_{mn}^{s}(k_{x})=\ell _{T}\int_{0}^{\infty }dx\
e^{-x}\;L_{m}(x)\int_{0}^{\infty }dx^{\prime }\ [K_{0}(|k_{x}\ell
_{T}||x-x^{\prime }|)+K_{0}(|k_{x}\ell _{T}||x+x^{\prime }|)]\ \tilde{R}%
_{0}(x^{\prime })\;e^{-x^{\prime }}\;L_{n}(x^{\prime }),  \label{16}
\end{equation}
and

\begin{eqnarray}
g_{mn}^{s}(k_{x}) &=&|k_{x}|\ell _{T}^{2}\int_{0}^{\infty }dx\
e^{-x}\{e^{-x/2}\;L_{m}(x)/\cosh (x/2)-\frac{m}{x}[L_{m}(x)-L_{m-1}(x)]\} 
\nonumber \\
* \   \nonumber \\
&&\int_{0}^{\infty }dx^{\prime }\ [\text{sign}\{x-x^{\prime
}\}K_{1}(|k_{x}|\ell _{T}|x-x^{\prime }|)+ K_{1}(|k_{x}|\ell
_{T}(x+x^{\prime}))]\ \tilde{R}_{0}(x^{\prime })\;e^{-x^{\prime
}}\;L_{n}(x^{\prime }).  \label{17}
\end{eqnarray}
Here sign$\{x\}=1$ for $x>0$ and sign$\{x\}=-1$ for $x<0$, and $K_{1}(x)$ is
the modified Bessel function. Notice that $r_{mn}^{s} \neq r_{nm}^{s}$ and $%
g_{mn}^{s} \neq g_{nm}^{s}$.

Now we consider the antisymmetric EMPs. From Eq. (\ref{7}), we obtain the
following integral equation for $\rho _{a}(\omega ,k_{x},\bar{y})$

\begin{eqnarray}
(\omega -k_{x}v_{g})&&\rho _{a}(\omega ,k_{x},\bar{y})- \frac{2}{\epsilon }%
\{(k_{x}\sigma _{yx}^{0}-ik_{x}^{2} \tilde{\sigma}_{xx})R_{0}(\bar{y})+i%
\tilde{\sigma}_{yy} \frac{d}{d\bar{y}}[R_{0}(\bar{y})\frac{d}{d\bar{y}}]\} 
\nonumber \\
* \   \nonumber \\
&&\int_{0}^{\infty }d\bar{y}^{\prime } [K_{0}(|k_{x}||\bar{y}-\bar{y}%
^{\prime}|)- K_{0}(|k_{x}||\bar{y}+\bar{y}^{\prime }|)] \rho _{a}(\omega
,k_{x},\bar{y}^{\prime })=0,  \label{12}
\end{eqnarray}
for $\bar{y} \geq 0$. As before, we write the exact solution of Eq. (\ref{12}%
) for the antisymmetric EMPs as

\begin{equation}
\rho _{a}(\omega ,k_{x},\bar{y})=\tilde{R}_{0}(Y)e^{-Y}\sum_{n=0}^{\infty
}\rho _{a}^{(n)}(\omega ,k_{x})L_{n}(Y),  \label{13}
\end{equation}
which must satisfy physically obvious boundary condition $\rho
_{a}(\omega,k_{x},0)=0$, which insures continuity of the $\rho _{a}(\omega
,k_{x},\bar{y})$ in the vicinity of $\bar{y}=0$; i.e., both for $\bar{y}%
\rightarrow +0$ and $\bar{y}\rightarrow -0$. The odd parity of $\rho
_{a}(\omega ,k_{x},\bar{y})$ imposes the following condition on $\rho
_{a}^{(n)}(\omega ,k_{x})$

\begin{equation}
\sum_{n=0}^{\infty }\rho _{a}^{(n)}(\omega ,k_{x})=0.  \label{14}
\end{equation}
For $\bar{y}<0,$ $\rho _{a}(\omega ,k_{x},\bar{y})$ can be obtained from Eq.
(\ref{13}) using the property $\rho _{a}(\omega ,k_{x},x)=-\rho
_{a}(\omega,k_{x},-x)$.

To obtain the dispersion equations for antisymmetric modes, we multiply Eq. (%
\ref{12}) by $L_{m}(Y)\tilde{R}_{0}^{-1}(Y)$ and integrate over $Y=\bar{y}%
/\ell _{T}$ from $0$ to $\infty $. Then, taking into account the Eq. (\ref
{13}), we obtain

\begin{equation}
(\omega -k_{x}v_{g})\rho _{a}^{(m)}(\omega ,k_{x})-
\sum_{n=0}^{\infty}[S\;r_{mn}^{a}(k_{x})+ S^{\prime
}\;g_{mn}^{a}(k_{x})]\rho _{a}^{(n)}(\omega,k_{x})=0,  \label{18}
\end{equation}
where

\begin{equation}
r_{mn}^{a}(k_{x})=\ell _{T}\int_{0}^{\infty }dx\
e^{-x}\;L_{m}(x)\int_{0}^{\infty }dx^{\prime }\
[K_{0}(|k_{x}|\ell_{T}|x-x^{\prime }|)- K_{0}(|k_{x}|\ell _{T}|x+x^{\prime
}|)]\ \tilde{R}_{0}(x^{\prime })\;e^{-x^{\prime }}\;L_{n}(x^{\prime }),
\label{19}
\end{equation}
and

\begin{eqnarray}
g_{mn}^{a}(k_{x}) &=&|k_{x}|\ell _{T}^{2}\int_{0}^{\infty }dx\
e^{-x}\{e^{-x/2}\;L_{m}(x)/\cosh (x/2)-\frac{m}{x}[L_{m}(x)-L_{m-1}(x)]\} 
\nonumber \\
* \   \nonumber \\
&&\int_{0}^{\infty }dx^{\prime }\ [\text{sign}\{x-x^{\prime
}\}K_{1}(|k_{x}|\ell _{T}|x-x^{\prime }|)-K_{1}(|k_{x}|\ell _{T}(x+x^{\prime
}))]\ \tilde{R}_{0}(x^{\prime })\;e^{-x^{\prime }}\;L_{n}(x^{\prime })+ 
\nonumber \\
\   \nonumber \\
&&2|k_{x}|\ell _{T}^{2}\int_{0}^{\infty }dx\ K_{1}(|k_{x}|\ell _{T}x)\; 
\tilde{R}_{0}(x)\;e^{-x}\;L_{n}(x).  \label{20}
\end{eqnarray}
In addition to above equations we must consider the condition given by Eq. (%
\ref{14}) that is essential to eliminate the logarithmic divergence in the
last integral in the RHS of Eq. (\ref{20}) after the pertinent summation
over $n$ in Eq. (\ref{18}). We point out that $r_{mn}^{a}\neq r_{nm}^{a}$, $%
g_{mn}^{a}\neq g_{nm}^{a}$ and hereafter we write $r_{mn}^{s,a}(k_{x})\equiv
r_{mn}^{s,a}$ and $g_{mn}^{s,a}(k_{x})\equiv g_{mn}^{s,a}$ to simplify the
notation.

In our solution of Eqs. (\ref{15})-(\ref{20}), we are taking the
long-wavelength limit $|k_{x}|\ell _{T}\ll 1$, so we can use the
approximations $K_{0}(|k_{x}|\ell _{T}x)\approx \ln (2/|k_{x}|\ell
_{T})-\gamma -\ln (x)$ and $K_{1}(|k_{x}|\ell _{T}x)\approx (|k_{x}|\ell
_{T}x)^{-1}$, where $\gamma $ is the Euler constant.

\section{SYMMETRIC AND ANTISYMMETRIC EMPs AT NOT-TOO-LOW TEMPERATURES}

\subsection{Symmetric modes}

Considering only the term $n=0$ in the RHS of Eq. (\ref{11}) and for $m=0$
in Eq. (\ref{15}), we obtain

\begin{equation}
\lbrack (\omega -k_{x}v_{g})-S\,r_{00}^{s}-S^{\prime }\,g_{00}^{s}]\rho
_{s}^{(0)}(\omega ,k_{x})=0,  \label{21}
\end{equation}
where

\begin{equation}
r_{00}^{s}=\ln (\frac{2}{|k_{x}|\ell _{T}})-\gamma -\frac{1}{4}
\int_{0}^{\infty }dx\ e^{-x}\;\int_{0}^{\infty }dx^{\prime }\ \frac{%
\ln(|x-x^{\prime }|)+\ln (x+x^{\prime })}{\cosh ^{2}(x^{\prime }/2)},
\label{22}
\end{equation}
and a numerical evaluation gives $g_{00}^{s}\approx 0.120$. After performing
the integrals In Eq. (\ref{22}), we have $r_{00}^{s}= \ln
(1/|k_{x}|\ell_{T})-0.012\approx \ln (1/|k_{x}|\ell _{T})$. Then the
dispersion relation (DR) can be written as

\begin{eqnarray}
\omega &=&k_{x}v_{g}+\ln (\frac{1}{|k_{x}|\ell _{T}})\;S+0.12S^{\prime } 
\nonumber \\
\   \nonumber \\
&\approx&k_{x}v_{g}+\frac{2}{\epsilon } [k_{x}\sigma _{yx}^{0}\ln (\frac{1}{%
|k_{x}|\ell _{T}})- 0.12i\frac{\tilde{\sigma}_{yy}}{\ell _{T}^{2}}],
\label{23}
\end{eqnarray}
where a quadratic contribution $\propto k_{x}^{2}\tilde{\sigma}%
_{xx}\ln(1/|k_{x}|\ell_{T})$ to the damping was neglected because we are
considering the long-wavelength limit. Corresponding to the DR of Eq. (\ref
{23}), $\rho_{s}(\omega ,k_{x},\bar{y})$ behaves as $\cosh ^{-2}(Y/2)$,
i.e., without any node. As a consequence, this is the DR of the fundamental
EMP at not-too-low temperatures. Furthermore, from Eqs. (\ref{23}) and (\ref
{9}), it follows that the damping rate of the fundamental EMP in this
temperature regime is proportional to $T^{-1}$ which is essentially
different from that at low temperatures \cite{balev97}. For $\nu =2$ the DR
of fundamental EMP can be obtained from Eq. (\ref{23}) by taking $\sigma
_{yx}^{0}=e^{2}/\pi \hbar $ and $\tilde{\sigma}_{yy}=e^{2}\ell
_{0}^{2}c^{^{\prime }}k_{B}T/(2\pi ^{2}\hbar ^{4}v_{g}^{3})$. Notice that
for the same $E_{e}$ we have Re $\omega (\nu =2)/$Re $\omega (\nu =1)=2$,
because $\ell _{T}(\nu =2)=\ell _{T}(\nu =1)$ and $v_{g}(\nu =2)=2v_{g}(\nu
=1)$ in contrast with the low temperature case \cite{balev97}. Notice that
both expressions for Re $\omega $ and Im $\omega $, given by Eq. (\ref{23}),
are essentially different from the results of the Volkov-Mikhailov model 
\cite{volkov88}. Here the characteristic length of the EMP is $\ell _{T}$,
while the characteristic length of a charge for EMP in the cited model is $%
\sigma_{yy}^{bu}/k_{x}\sigma _{yx}^{bu}$, where $\sigma _{\mu \gamma }^{bu}$
are the components of the bulk conductivity tensor of the 2DES.

Corrections to the DR of fundamental EMP given by Eq. (\ref{23}) and the
additional {\it symmetric} branch are obtained by keeping only the terms $%
n=0 $ and $n=1$ in Eq. (\ref{11}) which gives

\begin{equation}
\tilde{\rho}_{s}(\omega ,k_{x},Y)=\frac{1}{\cosh ^{2}(Y/2)}[1+\frac{\rho
_{s}^{(1)}(\omega ,k_{x})}{\rho _{s}^{(0)}(\omega ,k_{x})}L_{1}(Y)],
\label{24}
\end{equation}
where $\tilde{\rho}_{s}(\omega ,k_{x},Y)=4\ell _{T}\rho _{s}(\omega ,k_{x},%
\bar{y})/\rho _{s}^{(0)}(\omega ,k_{x})$. From Eq. (\ref{15}), for $m=0$, we
obtain

\begin{equation}
[(\omega -k_{x}v_{g})-S\,r_{00}^{s}-S^{\prime }\,g_{00}^{s}]\rho
_{s}^{(0)}(\omega ,k_{x})-[S\,r_{01}^{s}+S^{\prime }\,g_{01}^{s}]\rho
_{s}^{(1)}(\omega ,k_{x})=0,  \label{25}
\end{equation}
and for $m=1$

\begin{equation}
\lbrack (\omega -k_{x}v_{g})-S\,r_{11}^{s}-S^{\prime }\,g_{11}^{s}]\rho
_{s}^{(1)}(\omega ,k_{x})-[S\,r_{10}^{s}+S^{\prime }\,g_{10}^{s}]\rho
_{s}^{(0)}(\omega ,k_{x})=0,  \label{26}
\end{equation}
The solution of the determinantal equation of the above system yields two
branches $\omega _{+}^{s}(k_{x})$ and $\omega _{-}^{s}(k_{x})$. For $%
|k_{x}|\ell _{T}\ll 1,$ the numerical evaluation gives $r_{10}^{s}=0.347$, $%
r_{11}^{s}=0.240$, $g_{01}^{s}=0.215$, $g_{10}^{s}=0.261$, $g_{11}^{s}=0.406$%
. If we neglect the coupling terms, by formally setting $%
r_{01}^{s}=r_{10}^{s}=0$ and $g_{01}^{s}=g_{10}^{s}=0$, the Eq. (\ref{25})
gives the DR of the fundamental EMP, Eq. (\ref{23}), and, from the Eq. (\ref
{26}), we obtain the DR of the quadrupole EMP, {\it i.e}., the EMP with two
nodes at $\bar{y}=\pm \ell _{T}$, as

\begin{eqnarray}
\omega &=&k_{x}v_{g}+r_{11}^{s}\;S+g_{11}^{s}S^{\prime }  \nonumber \\
* &\approx&k_{x}v_{g}+\frac{2}{\epsilon }[0.24k_{x}\sigma_{yx}^{0}- 0.406i%
\frac{\tilde{\sigma}_{yy}}{\ell _{T}^{2}}].  \label{27}
\end{eqnarray}
The Re $\omega $ of the quadrupole EMP at not-too-low temperatures is almost
the same as that at low temperatures (cf. Eq. (24) of Ref. \cite{balev97}),
but is essentially different from the frequency of the quadrupole $j=2$
branch of the Aleiner-Glazman model \cite{aleiner94}.

For coupled modes the two branches are given by

\begin{eqnarray}
\omega _{\pm }^{s} &=&k_{x}v_{g}+\frac{1}{2}[S(r_{00}^{s}+r_{11}^{s})+
S^{\prime }(g_{00}^{s}+g_{11}^{s})]\pm \frac{1}{2}
\{[S(r_{00}^{s}-r_{11}^{s})+  \nonumber \\
&&S^{\prime }(g_{00}^{s}-g_{11}^{s})]^{2}+4(Sr_{01}^{s}+S^{\prime
}g_{01}^{s})(Sr_{10}^{s}+S^{\prime }g_{10}^{s})\}^{1/2}.  \label{28}
\end{eqnarray}
If not stated otherwise, we consider not too strong dissipation, for which $%
S\ln (1/|k_{x}|\ell _{T})\gg |S^{\prime }|$. In the long wavelength limit,
we obtain

\begin{equation}
\omega _{+}^{s}=k_{x}v_{g}+Sr_{00}^{s}+S^{\prime }g_{00}^{s}+\frac{%
(Sr_{01}^{s}+S^{\prime }g_{01}^{s})(Sr_{10}^{s}+S^{\prime }g_{10}^{s})}{%
S(r_{00}^{s}-r_{11}^{s})+S^{\prime }(g_{00}^{s}-g_{11}^{s})},  \label{29}
\end{equation}
and

\begin{equation}
\omega _{-}^{s}=k_{x}v_{g}+Sr_{11}^{s}+S^{\prime }g_{11}^{s}-\frac{%
(Sr_{01}^{s}+S^{\prime }g_{01}^{s})(Sr_{10}^{s}+S^{\prime }g_{10}^{s})}{%
S(r_{00}^{s}-r_{11}^{s})+S^{\prime }(g_{00}^{s}-g_{11}^{s})}.  \label{30}
\end{equation}
Further, after substituting the coefficients in Eq. (\ref{29}), we can write

\begin{equation}
\omega _{+}^{s}\approx k_{x}v_{g}+Sr_{00}^{s} [1-1/(7.5\ln
(1/k_{x}\ell_{T}))]+S^{\prime }g_{00}^{s}[1-10/12].  \label{31}
\end{equation}
We observe that by taking into account the coupling between the monopole and
quadrupole terms Re $\omega _{+}^{s}$ becomes slightly smaller. However, the
damping rate of the fundamental branch is decreased by 6. Finally, we obtain

\begin{equation}
\omega _{+}^{s}=k_{x}v_{g}+\frac{2}{\epsilon }[k_{x}\sigma _{yx}^{0} \ln (%
\frac{1}{|k_{x}|\ell _{T}})-i\frac{\tilde{\sigma}_{yy}} {\ell _{T}^{2}}%
(0.02+k_{x}^{2}\ell _{T}^{2} \ln(\frac{1}{|k_{x}|\ell _{T}}))],  \label{31a}
\end{equation}
where in the damping rate the term $\propto k_{x}^{2}$ is returned due to
essential suppression of the main contribution to Im$\omega_{+}^{s}$.
Substituting the DR (\ref{31}) in Eq. (\ref{26}) we obtain $\rho
_{s}^{(1)}(\omega ,k_{x})/\rho _{s}^{(0)}(\omega ,k_{x})\approx
(r_{10}^{s}/r_{00}^{s})(1+0.75S^{\prime }/S)\ll 1$, proving the fast
convergence of the expansion for this mode. Now, from the Eq. (\ref{24}),
the charge density for the renormalized fundamental EMP, for $Y\geq 0$, can
be written as

\begin{equation}
\tilde{\rho}_{s}(\omega _{+}^{s},k_{x},Y)=\frac{1}{\cosh ^{2}(Y/2)}[1-i\frac{%
0.26\tilde{\sigma}_{yy}}{\sigma _{yx}^{0}k_{x}\ell _{T}^{2}\ln
(1/|k_{x}|\ell _{T})}L_{1}(Y)].  \label{32}
\end{equation}
We observe that if for some phase $\phi $ of the wave, its amplitude along $%
y $ has a pure monopole character $\propto \cosh ^{-2}(Y/2)$ {\it i.e}.
without any node, after a shift of $\pm \pi /2$ in $\phi $, it acquires a
pure quadrupole character $\propto L_{1}(|Y|)\cosh ^{-2}(Y/2)$ with two
nodes at $Y=\pm 1$. Because such a behavior can be seen as the rotation of a
complex vector function while the wave propagates, we call this fundamental
EMP, characterized by the Eqs. (\ref{31}) and (\ref{32}), the {\it edge
helicon} (EH) for not-too-low temperatures which has different properties
from its counterpart at low temperatures \cite{balev97}.

From Eq. (\ref{30}) and after substituting the coefficients, the DR of the
other branch is given by 
\begin{equation}
\omega _{-}^{s}\approx k_{x}v_{g}+Sr_{11}^{s}[1+0.56]+S^{\prime
}g_{11}^{s}[1+1/4].  \label{33}
\end{equation}
Now the coupling between quadrupole and monopole terms leads to an increase
of $50\%$ in Re $\omega _{-}^{s}$ and $25\%$ in the damping rate. The ratio
between the amplitudes is $\rho _{s}^{(1)}(\omega ,k_{x})/\rho
_{s}^{(0)}(\omega ,k_{x})\approx 1/[2\ln (2)-1]\approx 2.6$ which gives a
rather small value for $\rho _{s}^{(0)}/\rho _{s}^{(1)}\approx 1/2.6$ and
the convergence of the expansion for the quadrupole mode. Then for
renormalized quadrupole mode $\tilde{\rho}_{s}(\omega ,k_{x},Y)$ is given,
from Eq. (\ref{24}), for $Y\geq 0$ as

\begin{equation}
\tilde{\rho}_{s}(\omega _{-}^{s},k_{x},Y)=\frac{1}{\cosh ^{2}(Y/2)}\left\{
1+[2\ln (2)-1]^{-1}L_{1}(Y)\right\} .  \label{34}
\end{equation}
Due to the coupling between the quadrupole and monopole modes, the nodes of $%
\tilde{\rho}_{s}(\omega ,k_{x},Y)$ are shifted, in the case of the
renormalized quadrupole mode, to $Y=\pm 2\ln (2)\approx \pm 1.39$. Notice
that $\rho _{s}(Y)$ for both $Y>0$ and $Y<0$ is given also by Eqs. (\ref{32}%
), (\ref{34}) after the replacement of $Y$ by $|Y|$.

In Fig. 2 the charge density profiles $\rho (Y)$ of EMPs are depicted for $%
\nu =2$ and $B=5.9$ T. The parameters are the same as in Fig. 1 and, when
necessary, we took $k_{x}\ell _{T}=0.1$ and $\epsilon=12.5$. The curves
labeled 1 and 2 represent $\rho (Y)\equiv \rho _{s}(Y,k_{x})=$ Re $[\tilde{%
\rho}_{s} (\omega_{+}^{s},k_{x},|Y|)\times \exp (i\phi )]$ for the EH, given
by Eq. (\ref{32}), if the wave phase $\phi =2\pi N$ and $\phi =\pi /2+2\pi N$%
, where $N$ is integer, respectively. The curve 3 represents $\rho (Y)=%
\tilde{\rho}_{s}(\omega _{-}^{s},k_{x},|Y|)$ for the renormalized quadrupole
mode, given by Eq. (\ref{34}).

\subsection{Antisymmetric modes}

Similar analysis can be done for studying the antisymmetric modes. The DR of
the dipole mode at not-too-low temperatures, after taking into account only
the terms $n=0$ and $n=1$ in the RHS of Eq. (\ref{13}), is given by 
\begin{eqnarray}
\omega&=&k_{x}v_{g}+(r_{00}^{a}-r_{01}^{a})\;
S+(g_{00}^{a}-g_{01}^{a})S^{\prime }  \nonumber \\
* \   \nonumber \\
&\approx& k_{x}v_{g}+\frac{2}{\epsilon } [0.65\;k_{x}\sigma _{yx}^{0}-0.5\;i 
\frac{\tilde{\sigma}_{yy}}{\ell _{T}^{2}}],  \label{36}
\end{eqnarray}
since $r_{00}^{a}\approx 0.509$, $r_{01}^{a}\approx -0.141$ and $%
g_{00}^{a}-g_{01}^{a}-1\approx -0.502$. The charge amplitude $%
\rho_{a}(\omega ,k_{x},\bar{y})$ of this dipole mode behaves as $\tilde{\rho}%
_{a}(Y)= 4 \ell_{T} \rho_{a}(\omega,k_{x},\bar{y})/ \rho^{(1)}_{s}(\omega,
k_{x})=-Y \cosh^{-2}(Y/2)$, {\it i.e}., with one node at $Y=0$. Its density
profile $\rho (Y)\equiv \tilde{\rho}_{a}(Y)$ corresponds to curve 4 in Fig.
2 . The frequency of the dipole EMP, Re $(\omega -k_{x}v_{g})/S=0.65,$ is $%
30\%$ larger than for the similar mode at low temperatures.

Corrections to the DR of the dipole EMP, given by Eq. (\ref{36}), and the
additional {\it antisymmetric} branch can also be obtained, by keeping only
the terms $n=0,1$ and $n=2$ in the RHS of Eq. (\ref{13}), in a
straightforward way. Then we obtain two branches $\omega_{+}^{a}(k_{x})$ and 
$\omega _{-}^{a}(k_{x})$. If we neglect the coupling between different terms
of the expansion we obtain, for $|k_{x}|\ell _{T} \ll 1$, the following
expression for the DR of the octupole mode

\begin{equation}
\omega \approx k_{x}v_{g}+\frac{2}{\epsilon }[0.14k_{x}\sigma
_{yx}^{0}-0.205i\frac{\tilde{\sigma}_{yy}}{\ell _{T}^{2}}].  \label{39}
\end{equation}
Neglecting dissipation and the $k_{x}v_{g}$ term, the octupole EMP at
not-too-low temperatures exhibits a phase velocity around $85\%$ of the
phase velocity of the octupole EMP at low temperatures (cf. Eq. (44) of Ref. 
\cite{balev97}). Again the frequency of this mode is distinct from the
frequency of the octupole $j=3$ branch of Ref. \cite{aleiner94}. The charge
density profile $\tilde{\rho}_{a}(Y)=Y(1-|Y|/2)\,\cosh ^{-2}(Y/2)$, {\it i.e}%
., with three nodes at $Y=0$ and $Y=\pm 2$ is represented by curve 5 in Fig.
2.

If we consider the coupling between the expansion terms, we obtain the
dipole and octupole renormalized modes. Our numerical results have shown
that the renormalization effects are quite weak in the regime of not-too-low
temperatures which means that, in our theory for high temperature EMPs, the
convergence of the expansion, given by Eq. (\ref{13}), for the dipole and
octupole modes is faster than that expansion over Hermitian polynomials used
in Ref. \cite{balev97} in the regime of very low temperatures.

In Fig. 3 we plot the dimensionless electric potential $\Phi \propto
\int_{-\infty }^{\infty }dY^{\prime }\ K_{0}(|k_{x}|\ell _{T}|Y-Y^{\prime
}|)\ \rho (Y^{\prime })$ for EMPs with the charge density profiles $\rho (Y)$
denoted by curves 1 to 5 in Fig. 2.

\section{DISCUSSION AND CONCLUDING REMARKS}

In this work we have introduced an analytical unperturbed electron density
profile $n_{0}(Y)/n_{0}=[1-\tanh (Y/2)]/2$ which is in very good agreement
with the exact results, as shown in Fig. 1, when the conditions, $\hbar
\omega _{c}/k_{B}T\gg 1$, $k_{e}\gg k_{B}T/\hbar v_{g}\gg \ell _{0}^{-1}$
are fulfilled.

We have shown that temperature effects manifest themselves in the dispersion
and spatial structure of the EMP modes by changing the characteristic length
to $\ell _{T}\gg \ell _{0}$. The edge density profiles for dipole,
quadrupole, and octupole EMPs in this regime behave spatially independent of
the wave phase $\phi $. However the behavior of the density profile of the
edge helicon is qualitatively modified by varying $\phi $ as shown in curves
1 and 2 of Fig. 2. Concerning the dissipation of the modes, we introduce the
dimensionless parameter $\eta _{T}$ as $\eta _{T}=\xi /(k_{x}\ell _{T})$,
where $\xi =\tilde{\sigma}_{yy}/(\ell _{T}\sigma
_{yx}^{0})=[e^{2}h_{14}^{2}/4\pi \hbar s^{3}\rho _{V}](s^{2}/v_{g}^{2})$.
Then, we have observed that the regime of {\it weak dissipation }occurs at $%
\eta _{T}\ll 1$ since Re $\omega \gg $ Im $\omega $ for all modes. The
opposite regime of {\it strong dissipation} corresponds to $\ln (1/k_{x}\ell
_{T})\gg \eta _{T}\agt 1$ because all modes, except the edge helicon, are
strongly damped, {\it i.e}., Re $\omega \alt$ Im $\omega $. So the edge
helicon is the only weakly damped mode in the region of strong dissipation.
We also observe from the comparison between the curves in Fig. 3 and the
corresponding ones in Fig. 2 that the wave potential $\Phi (Y)\propto
E_{x}(\omega ,k_{x},y)$ is smooth on the scale of the magnetic length both
for symmetric and antisymmetric modes even when the charge density profile $%
\rho (Y)$ of the EMP shows a cusp in the vicinity of $Y=0$. This smoothness
of the EMPs electric potential justifies well the assumptions of our present
study.

Finally, we make some estimates that should be useful in experimental
studies. For the GaAs-based samples with the parameters used in Fig. 1 and $%
k_{x}\ell _{T}=0.1$ ($k_{x}\approx 2\times 10^{4}$ cm$^{-1}$), we obtain $%
\xi=0.353s^{2}/v_{g}^{2}\approx 8.8\times 10^{-2}$ and hence $\eta _{T}
\approx 0.88$, which corresponds to the case of {\it strong dissipation}. We
propose here a sample arrangement very similar to that of Ashoori {\it et. al%
}.\cite{ashoori92}. Our circular mesa has a diameter $D=15\ \mu $m (in Ref. 
\cite{ashoori92} $D=540\ \mu $m) and height $\sim 1\mu $m in the middle of a
wide GaAs/AlGaAs chip with large thickness such that $2d_{s}\gg 1/k_{x}=0.5\
\mu $m and we assume as in Ref. \cite{ashoori92} the value $d_{s}=500\ \mu $%
m. The condition $2|k_{x}|d_{s}\gg 1$ is well satisfied even for $%
k_{x0}=2/D=1.3\times 10^{3}$ cm$^{-1}$, as the assumed $k_{x} \gg k_{x0}$.
As a consequence, the conditions $2k_{x}d_{s}\gg 1$ and $k_{x}W \gg 1 $ are
satisfied as well. Analogous to Ref. \cite{ashoori92}, we assume that a
square ``pulser'' gate with width $L_{p}\approx 0.8\ \mu $m ($L_{p}=10\mu $m
in Ref. \cite{ashoori92}) is much smaller than $\pi D$. Furthermore, the
initial charge distribution (when a pulse of external voltage is applied to
the ''pulser'' gate) has a rectangular form and therefore the essential
contributions comes from the $k_{xn}=\pm nk_{x0},\ n=1,2,..,$ modes
distributed in the interval $k_{x0}\leq |k_{xn}|<\pi /L_{p}$. So, a typical
wave number $k_{xt}=|k_{xnt}|\approx \pi /2L_{p}=2\times 10^{4}$ cm$^{-1}$
is of the same order of the magnitude of our estimated $k_{x}$. As all
modes, except one, are strongly damped we are left with the edge helicons.
Equation (\ref{31a}) gives a decay rate Im $\omega _{+}^{s} \approx
7.5\times 10^{9}$ sec$^{-1}$. The corresponding group velocity for this mode
is $v_{g+}(k_{xt})=v_{g}+(2/\epsilon )\sigma _{yx}^{0} [ln(1/k_{xt}
\ell_{T})-1]$ and gives a period $T_{+}=\pi D/v_{g+}(k_{xt}) \approx 3.1
\times 10^{-10}$ sec. So, during the travelling period $T_{+},$ the
amplitude of the mode should drop only by a factor $\sim exp(-2.3)$. As it
is known, the amplitude of the travelling pulse in time-resolved experiments%
\cite{ashoori92} can be measured by another square gate with side $L_{p}$
above the edge. So, we believe from our reasonable estimates that the edge
helicon mode may be detected.

\acknowledgements

This work was supported by Brazilian FAPESP Grants No.98/10192-2 and
95/0789-3. In addition, O. G. B. acknowledges partial support by the
Ukrainian SFFI Grant No. 2.4/665, and N. S. is grateful do Brazilian CNPq
for a research fellowship.

\newpage%

\begin{center}
FIGURE CAPTIONS
\end{center}

\vspace{0.5cm}%

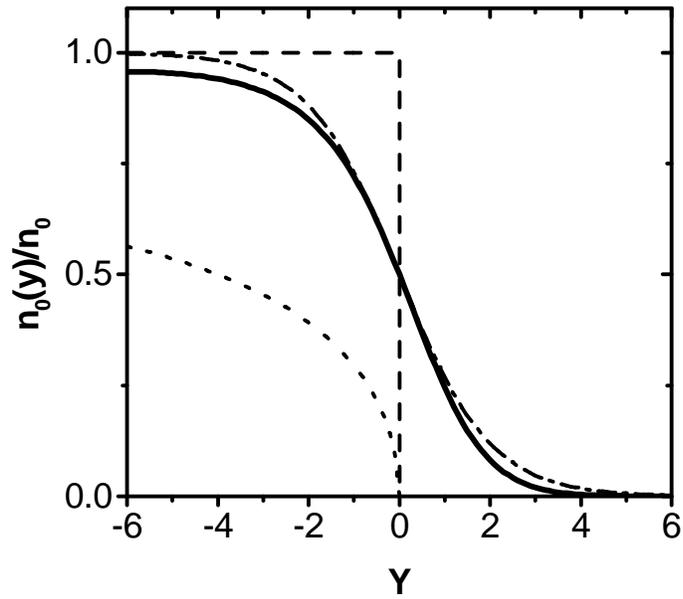
FIG. 1. Unperturbed electron density $n_{0}(y)$, normalized to the bulk
value $n_{0}$, as a function of $Y=(y-y_{r0})/\ell _{T}$, where $y_{r0}$ is
the edge of $n=0$ LL. The solid and dash-dotted curves show our exact and
approximate profiles, respectively, for a GaAs-based heterostructure and for 
$\nu =2$, $B=5.9$T, $\omega _{c}/\Omega =30$, $\Delta _{F}=\hbar \omega
_{c}/2$, $\ell _{T}/\ell _{0}=5$, and $T=18$ K. The dashed curve is the
density profile in the model of Ref. \cite{volkov88} and the dotted curve
that of Ref. \cite{aleiner94} for $n_{0}(y)/n_{0}=(2/\pi )\arctan
[(y_{re}-y)/a]^{1/2},$ $a/\ell _{0}=20$. For the dashed and the dotted
curves $Y=(y-y_{re})/\ell _{T}$, where $y_{re}$ is the edge in the models of
Refs. \cite{volkov88} and \cite{aleiner94}.

\vspace{0.7cm}%

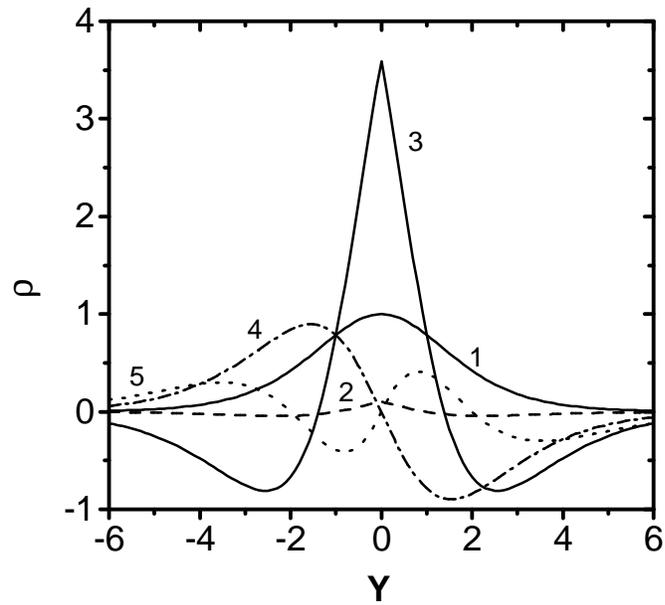
FIG. 2. Dimensionless charge density profile $\rho (Y)$ of EMPs at
not-too-low temperatures: curves 1, 2, and 3 correspond to symmetric modes
and curves 4 and 5 to the antisymmetric ones. Curves 1 and 2 represent $\rho
(Y)$ for the edge helicon, using Eq. (\ref{32}), for different wave phases
and curve 3 represents $\rho (Y)$ for the renormalized quadrupole EMP, Eq. (%
\ref{34}). Curves 4 and 5 are $\rho (Y)$ for dipole and octupole EMPs,
respectively. The parameters are the same as in Fig. 1.

\vspace{0.7cm}%

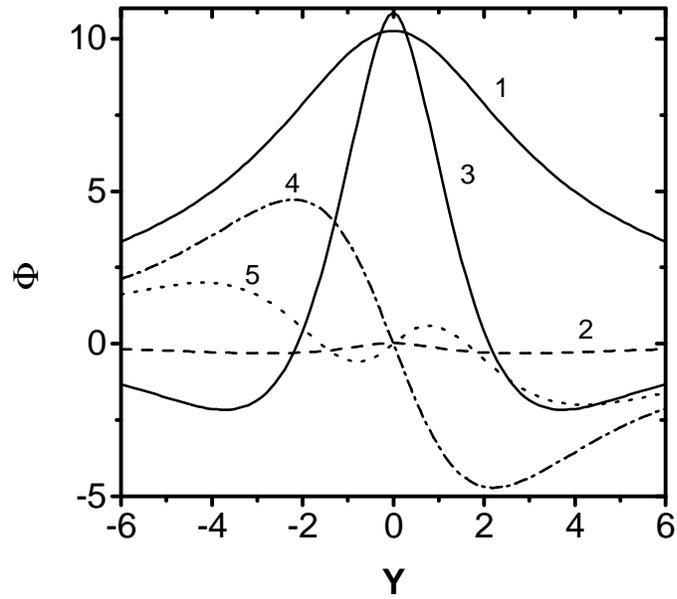
FIG. 3. Dimensionless electric potential $\Phi (Y)$ of EMPs at not-too-low
temperatures. Curves 1 to 5 correspond to charge profiles labeled in Fig. 2.

\end{document}